\documentclass[twocolumn,twoside]{IEEEtran/IEEEtran}
\usepackage{times,helvet}
\usepackage{amsfonts}
\usepackage{amsmath}
\usepackage{amssymb}
\usepackage{wrapfig,booktabs}
\usepackage{hyperref}
\usepackage{enumitem}
\usepackage{graphicx}
\usepackage{rotating}
\usepackage{ulem}
\usepackage[T1]{fontenc}
\usepackage{tikz}
\usetikzlibrary{shapes,arrows}

\definecolor{olive}{rgb}{0.3, 0.8, .1}
\definecolor{lightblue}{rgb}{0, 0.5, 1}

\newcommand{\sw}{\mathsf{w}}

\newcommand{\sy}{\mathsf{y}}
\newcommand{\sz}{\mathsf{z}}

\newcommand{\cordl}[1]{}

\newcommand{\engdl}[1]{}

\title{\LARGE \bf
KLTS: A rigorous method to compute the confidence intervals \\
for the Three-Cornered Hat and for Groslambert Covariance
}

\author{\'{E}ric Lantz$^{1\star}$, Claudio E. Calosso$^{2}$, Enrico Rubiola$^{3,2}$, Vincent Giordano$^{3}$, Christophe Fluhr$^{4}$, Benoît Dubois$^{4}$ and Fran\c{c}ois Vernotte$^{3\star}$
\thanks{$^{1}$D\'{e}partement d'Optique P.M. Duffieux, FEMTO-ST, UMR 6174 CNRS-Universit\'e Bourgogne Franche-Comt\'e, France}%
\thanks{$^{2}$Physics Metrology Division, Istituto Nazionale di Ricerca Metrologica
INRiM, Torino, Italy}%
\thanks{$^{3}$Department of Time and Frequency,
FEMTO-ST, Observatory THETA, UMR 6174 CNRS-Universit\'e Bourgogne Franche-Comt\'e, France}%
\thanks{$^{4}$FEMTO Engineering, Besançon, France}%
\thanks{$^\star$Corresponding authors: {\tt\small eric.lantz@univ-fcomte.fr}, {\tt\small francois.vernotte@obs-besancon.fr}}%
}

\begin{document}

\maketitle
\begin{abstract}

The three-cornered hat / Groslambert Covariance methods are widely used to estimate the stability of each individual clock in a set of three, but no method gives reliable confidence intervals for large integration times. 

We propose a new KLTS (Karhunen-Lo\`eve Tansform using Sufficient statistics) method which uses these  estimators to take into account the statistics of all the measurements between the pairs of clocks in a Bayesian way. The resulting Cumulative Density Function (CDF) yields confidence intervals for each clock AVAR. This CDF provides also a stability estimator which is always positive. 

Checked by massive Monte-Carlo simulations, KLTS proves to be perfectly reliable even for one degree of freedom. An example of experimental measurement is given.
\end{abstract}





\section{INTRODUCTION}

Although the three-cornered hat \cite{gray1974} and the Groslambert Covariance (GCov) \cite{groslambert1981} methods are widely used to measure the stability of each individual clock in a set of three, the only methods which exist to compute error bars are limited to the smallest integrations times, i.e. when the number of Equivalent Degrees of Freedom (EDF) is high \cite{ekstrom2006, vernotte2016, vernotte2018}.
However, there is no reliable method to assess confidence intervals over the estimates if their number of EDF is low, typically 5 or below. 
 However, since this case occurs for the largest integration times, it is an important issue for all applications dealing with long term stability (e.g. time keeping). 

Likewise, another problem frequently arises using the three-cornered hat or the GCov method: negative variance estimates can occur. Although this issue was already considered by Premoli and Tavella \cite{premoli1993}, it would be useful to get a method which could provide simultaneously a positive estimate as well as its confidence interval.

In a previous paper, we performed a first Bayesian attempt to estimate confidence intervals from the statistics of the three-cornered estimates but we observed that this method was only valid beyond 5 EDF \cite{vernotte2018}. We propose here a new method, also based on Bayesian inversion, which uses the statistics of the data themselves, instead of the statistics of the estimates. Nevertheless, these statistics of the data can be computed from the values of the estimates, since these estimates form a ``sufficient statistics'' \cite{kholevo2001} for the variance estimation. The resulting Cumulative Density Functions (CDF) yield the lower and upper bounds of the 95 \% confidence interval. On the other hand, the CDF at 50 \% provides a useful stability estimator of the stability of each clock, i.e. the median value, which has the advantage of always being positive. 

The performances of this method have been checked by using massive Monte-Carlo simulations. The principle of these simulations is described in this paper and the comparisons with the theoretical confidence intervals given by our new method are discussed. Finally, this method is applied to the measurement of 3 Cryogenic Sapphire Oscillators (CSO). 

\section{Statement of the Problem}
\subsection{Clock comparisons}
\begin{figure}
\centering
\begin{tikzpicture}
\begin{scope}[scale=0.4]
\fill[blue!15] (0,0) circle(2);
\draw[line width=1.5pt] (0,0) circle(2);
\draw (0,0) node{\huge$\sim$};
\draw (2,0)--(4,0);
\draw (8,0)--(10,0);
\fill[blue!50] (4,-1)--(4,1)--(8,1)--(8,-1)--(4,-1);
\draw[line width=1.5pt] (4,-1)--(4,1)--(8,1)--(8,-1)--(4,-1);
\draw (0,-2.5) node{Clock $B$};
\draw (6,-1.5) node{Counter $1$};
\draw (6,0) node{TIC};

\fill[blue!15] (12,0) circle(2);
\draw[line width=1.5pt] (12,0) circle(2);
\draw (12,0) node{\huge$\sim$};
\draw (1.0000,1.7321)--(2.0000,3.4641);
\draw (4.0000,6.9282)--(5.0000,8.6603);
\fill[blue!50] (2.8660,2.9641)--(1.1340,3.9641)--(3.1340,7.4282)--(4.8660,6.4282)--(2.8660,2.9641);
\draw[line width=1.5pt] (2.8660,2.9641)--(1.1340,3.9641)--(3.1340,7.4282)--(4.8660,6.4282)--(2.8660,2.9641);
\draw (6,12.892) node{Clock $A$};
\draw (1.7010,5.9462) node[rotate=60]{Counter $3$};
\draw (3,5.1962) node[rotate=60]{TIC};

\fill[blue!15] (6,10.392) circle(2);
\draw[line width=1.5pt] (6,10.392) circle(2);
\draw (6,10.392) node{\huge$\sim$};
\draw (11.0000,1.7321)--(10.0000,3.4641);
\draw (8.0000,6.9282)--(7.0000,8.6603);
\fill[blue!50] (9.1340,2.9641)--(10.8660,3.9641)--(8.8660,7.4282)--(7.1340,6.4282)--(9.1340,2.9641);
\draw[line width=1.5pt] (9.1340,2.9641)--(10.8660,3.9641)--(8.8660,7.4282)--(7.1340,6.4282)--(9.1340,2.9641);
\draw (12,-2.5) node{Clock $C$};
\draw (10.2990,5.9462) node[rotate=-60]{Counter $2$};
\draw (9,5.1962) node[rotate=-60]{TIC};
\end{scope}
\end{tikzpicture}
\caption{Layout of the clock measuring device.\label{fig:layout}}
\end{figure}
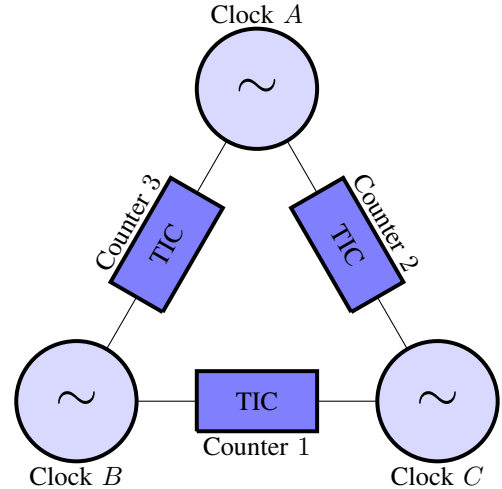
Let us consider 3 independent clocks: $A$, $B$ and $C$. It is
possible to compare these clocks by pairs by using 3 Time Interval Counter (TIC), as shown in Figure \ref{fig:layout}, and to estimate the
corresponding Allan variances (AVAR). Let us denote $\bar{\sy}_{ABk}$ the $k^\mathrm{\footnotesize th}$ frequency deviation sample between $A$ and $B$ integrated without dead time over a duration $\tau$, $\bar{\sy}_{ABk}=\frac{1}{\tau}\int_{t_k}^{t_k+\tau}\sy(t)\textrm{d}t$, and $\sz_{ABk}=\left(\bar{\sy}_{ABk+1}-\bar{\sy}_{ABk}\right)/\sqrt{2}$. For clocks $A$ and $B$, the AVAR is 
\begin{equation}
\sigma_{AB}^2(\tau)=\mathbb{E}\left[ \sz_{ABk}^2\right]
\label {varab}
\end{equation}
where $\mathbb{E}[\cdot]$ stands for the mathematical expectation of the quantity between the brackets. For the sake of concision of the notations, we will drop the dependence on $\tau$ in the following.
Since $\sigma_{AB}^2=\sigma_A^2+\sigma_B^2$, the
three-cornered hat is based on the following property:
\begin{equation}
\textrm{TCH}_A=\frac{1}{2}\left(\sigma_{AB}^2-\sigma_{BC}^2+\sigma_{CA}^2\right)=\sigma_A^2.\label{eq:theo_TCH}
\end{equation}

On the other hand, the Groslambert covariance is based on this
other property:
\begin{equation}
\textrm{GCov}_A=\mathbb{E}\left[ \sz_{ABk}\cdot\sz_{ACk}\right]=\sigma_A^2.\label{eq:theo_GCov}
\end{equation} 

	\subsection{Measurement noise influence}
Let us consider the layout of Figure \ref{fig:layout}:
\begin{itemize}
	\item the clocks $B$ and $C$ are connected to the counter $1$ which is affected by a measurement noise $\left\{\sz_{N1k}\right\}$
	\item the clocks $C$ and $A$ are connected to the counter $2$ which is affected by a measurement noise $\left\{\sz_{N2k}\right\}$
	\item the clocks $A$ and $B$ are connected to the counter $3$ which is affected by a measurement noise $\left\{\sz_{N3k}\right\}$.
\end{itemize}
We assume that these 3 measurement noises are uncorrelated.

The measurement given by the counter $3$ is then the sum of the clock noises plus the corresponding measurement noise: $\sz_{3k}=\sz_{ABk} + \sz_{N3k}$. Its variance is
\begin{equation}
 \sigma_3^2=\mathbb{E}\left[\left(\sz_{ABk} + \sz_{N3k}\right)^2\right]=\sigma_{AB}^2+\sigma_{N3}^2=\sigma_{A}^2+\sigma_{B}^2+\sigma_{N3}^2.
 \label {var3}
 \end{equation}
 In the following, we will call $\sz_{1k}$, $\sz_{2k}$ and $\sz_{3k}$ the \textbf{measurements}.

The three-cornered hat becomes $\textrm{TCH}_A=\frac{1}{2}\left(\sigma_{3}^2-\sigma_1^2+\sigma_2^2\right)=\sigma_A^2+\frac{1}{2}\left(\sigma_{N3}^2-\sigma_{N1}^2+\sigma_{N2}^2\right)$.
Let us assume that the 3 counter measurement noises are of the same level: $\sigma_{N1}^2\approx\sigma_{N2}^2\approx\sigma_{N3}^2=\sigma_{N}^2$. Therefore $\textrm{TCH}_A=\sigma_A^2+\frac{1}{2}\sigma_N^2$.

By contrast, the Groslambert covariance remains, because the measurement noises are uncorrelated:
\begin{equation}
 \textrm{GCov}_A=\mathbb{E}\left[ \left(\sz_{ABk} + \sz_{N3k}\right)\cdot\left(\sz_{ACk} + \sz_{N2k}\right)\right]=\sigma_A^2.
 \label {GCov}
\end{equation}
Therefore, the only difference between these two approaches concerns the measurement noise due to the counters, since the expectation of the GCov estimates is not sensitive to this noise \cite{vernotte2016}. Note however that the variance of the GCov estimates does include a measurement noise term. 

	\subsection{Model parameters and estimates}
The previously defined quantities $\sigma_A^2$, $\sigma_{AB}^2$, $\textrm{TCH}_A$, $\textrm{GCov}_A$, $\sigma_1^2$, $\ldots$ are unknown real values. However, they can be estimated by estimates which are random variables. 
 We can then define the following estimates:
\begin{eqnarray}
\hat{\sigma}_3^2&=&\frac{1}{M}\sum_{k=1}^M \sz_{3k}^2=\frac{1}{M}\sum_{k=1}^M \left(\sz_{ABk} + \sz_{N3k}\right)^2\nonumber\\
&=&\hat{\sigma}_A^2+\hat{\sigma}_B^2+\hat{\sigma}_N^2
\end{eqnarray}
where the hat ($\hat{\cdot}$) stands for \textit{estimate}, $M$ is the number of available consecutive $\sz_{ABk}$. Let us call the $\hat{\sigma}_1^2$, $\hat{\sigma}_2^2$ and $\hat{\sigma}_3^2$ the \textbf{elementary estimates}.

Similarly,
$\widehat{\textrm{TCH}}_A=\hat{\sigma}_A^2+\frac{1}{2}\hat{\sigma}_N^2$
and
\begin{eqnarray}
\widehat{\textrm{GCov}}_A&=&\frac{1}{M}\sum_{k=1}^M \sz_{3k}\sz_{2k}\nonumber\\
&=&\frac{1}{M}\sum_{k=1}^M \left(\sz_{ABk} + \sz_{N3k}\right)\left(\sz_{ACk} + \sz_{N2k}\right)\nonumber\\
&=&\hat{\sigma}_A^2.
\end{eqnarray} 
Let us call the $\hat{\sigma}_A^2$, $\hat{\sigma}_B^2$ and $\hat{\sigma}_C^2$ the \textbf{final estimates}.

Meanwhile, let us call the $\sigma_A^2$, $\sigma_B^2$ and $\sigma_C^2$ the \textbf{model parameters}.

The aim of this paper consists in calculating
a confidence interval over each model parameter $\sigma_A^2$, $\sigma_B^2$ and $\sigma_C^2$,
from the final and elementary estimates.

	\subsection{Estimation of the measurement noise by the closure relationship}
The closure relationship is obtained by computing the sum of the measurements of all counters for a given $k$:
\begin{eqnarray}
\sz_{cls,k}&=&\sz_{1k}+\sz_{2k}+\sz_{3k}\nonumber\\
&=&\sz_{BCk}+\sz_{N1k}+\sz_{CAk}+\sz_{N2k}+\sz_{ABk}+\sz_{N3k}\nonumber\\
&=&\sz_{N1k}+\sz_{N2k}+\sz_{N3k}.
\end{eqnarray}
since $\sz_{ABk}=\sz_{Bk}-\sz_{Ak}$.
These 3 measurement noises being uncorrelated and of equal level, the variance of the closure is: $\sigma_{cls}^2=\mathbb{E}\left[\left(\sz_{N1k}+\sz_{N2k}+\sz_{N3k}\right)^2\right]=3\sigma_N^2$.

This gives an efficient way to estimate the variance of the measurement noise:
\begin{equation}
\hat{\sigma}_N^2=\frac{1}{3M}\sum_{k=1}^M \left(\sz_{N1k}+\sz_{N2k}+\sz_{N3k}\right)^2=\frac{1}{3}\hat{\sigma}_{cls}^2.
\end{equation}

	\subsection{Bayesian inference}
In Bayesian analysis, we have to consider the model parameters $\vec{\Theta}=(\theta_1, \ldots, \theta_m)^T$ which have, in the model world,  $m$ definite but unknown values, and the measurements $\vec{X}=(x_1, \ldots, x_n)^T$ which are $n$ random variables. In our case, the parameters are the 3 true $\sigma_A^2$, $\sigma_B^2$ and $\sigma_C^2$ AVAR values of the 3 clocks and the measurements are either the elementary estimates $\hat{\sigma}_1^2$, $\hat{\sigma}_2^2$ and $\hat{\sigma}_3^2$ (see the previous method described in \cite{vernotte2018}) or directly the $3M$ measurements $\left\{\sz_{ABk},\sz_{BCk},\sz_{ACk}\right\}$. In the present method, we will compute the Probability Density Function (PDF) of these $3M$ measurements, by using only the measured values of the final and elementary estimates, because these estimates form a sufficient statistics for the measurements \cite{kholevo2001, saporta1990}: the precise knowledge of the set of measurements does not bring any new information beyond the estimates. 

While using the same  estimates as \cite{vernotte2018}, the present method computes only the gaussian PDF of the measurements themselves, while \cite{vernotte2018} lays on the approximation of the PDF of the estimators by a Gaussian law, inducing errors for a low number of EDF.

The Bayesian inversion lays on the distinction between two issues:
\begin{itemize}
	\item the \textbf{direct problem}, which consists, in the model world,  in calculating the Probability Density Function (PDF) of the estimates knowing the model parameters $p(\vec{X}|\vec{\Theta})$;
	\item the \textbf{inverse problem}, which consists, in the experimenter world, in calculating the PDF of the model parameters knowing the estimates $p(\vec{\Theta}|\vec{X})$. This is the most precise knowledge we can gain on these parameters after a measurement.
\end{itemize}
The direct problem has been solved in \cite{vernotte2018} and the inverse problem may be solved thanks to the Bayes theorem:
\begin{equation}
\left\{\begin{array}{l}
\displaystyle p(\vec{\Theta}|\vec{X})\propto \pi(\vec{\Theta}) \cdot p(\vec{X}|\vec{\Theta})\\
\displaystyle \int p(\vec{\Theta}|\vec{X}) \textrm{d}\vec{X}=1
\end{array}\right.\label{eq:bayes_inference}
\end{equation}
where $\pi(\vec{\Theta})$, called the prior, is the \textit{a priori} probability of the parameter $\vec{\Theta}$ before any measurement.

\section{The KLTS Method}
In the following, in order to distinguish our previous method described in \cite{vernotte2018} and the present method, we will call the former the KLTG method, for ``Karhunen-Lo\`eve Transform with Gaussian approximation'' method, and the latter the KLTS method, for ``Karhunen-Lo\`eve Transform using Sufficient statistics'' method.

	\subsection{Using the measurements $\sz$}

The KLTS method relies on the use of the $\sz_{1k}$, $\sz_{2k}$, $\sz_{3k}$ measurements, which are Gaussian random variables (r.v.), instead of the  $\hat{\sigma}_1^2$, $\hat{\sigma}_2^2$ and $\hat{\sigma}_3^2$ elementary estimates, which are a linear combination of $\chi^2$ random variable. The main advantage of this approach lays in the property of the Gaussian r.v. which remain Gaussian when they are linearly combined.

However, these measurements are strongly correlated for two reasons:
\begin{enumerate}
	\item the $\sz_{ABk}$ and $\sz_{ABk+1}$ are not independent (except in the case of White FM and AVAR without overlapping);
	\item for a given $k$, the $\sz_{1k}$, $\sz_{2k}$ and $\sz_{3k}$ are not independent (e.g. their sum is null if the measurement noise is neglected).
\end{enumerate}
We will postpone the treatment of the correlation between successive measurements. We first treat the correlations between the 3 measurements at a given time. Hence, let us assume that the $3M$ measurements form $M$ independent triplets, successive realizations of 3 correlated Gaussian r.v. We aim to determine three linear combinations of these r.v. that are independent one of each other:

\begin{equation}
\sw_{l,k} =\alpha_l\quad\!\!\!\! \sz_{1,k}+\beta_l\quad\!\!\!\!\sz_{2,k}+\gamma_l\quad\!\!\!\!\sz_{3,k},\quad\!\!\!\! l=1,2,3
\label {y}
\end{equation}

The solution of this problem is given by the Karhunen-Lo\`eve (K.L.) transform: the 9 coefficients $\alpha_l, \beta_l, \gamma_l$ form the eigenvectors of the rotation matrix which diagonalizes the covariance matrix of the measurements, whose diagonal elements are given by Eq.(\ref {var3}) and non diagonal elements by Eq.(\ref{GCov}). Note that this matrix is singular in the absence of measurement noise, because $\sz_{BCk}=\sz_{ACk}-\sz_{ABk}$: if the measurement noise is negligible, a 2 by 2 matrix must be used instead of a a 3 by 3 matrix. 

Now, since the $\sw_{l,k}$ are all independent, the PDF is easy to calculate in the model world, by assuming definite values for the true variances: 
$p(\vec{W}|\vec{\Theta})=\prod_{k=1}^{M}\prod_{l=1}^{3}p(\sw_{l,k}|\vec{\Theta})$. To render more explicit this expression, let us introduce the three K.L. variances $V_l$ obtained by diagonalization of the covariance matrix.  All PDF are Gaussian and their product can be written as:  
\begin{equation}
p(\vec{W}|\vec{\Theta})=\prod_{l=1}^{3}\frac{1}{V_l^{M/2}}\exp\left(-\frac{\sum_{k=1}^M \sw_{l,k}^2}{2V_l}\right)
\label{proba}
		\end{equation}

\subsection{Using the sufficient statistic properties of the estimates}

In this equation, it is quite interesting to develop, using Eq. (\ref {y}), the numerator of the exponential argument, which is the only term that depends on the actual measurements:

\begin{eqnarray}
\sum_{k=1}^M \sw_{l,k}^2&=&M \left[\alpha_l^2\hat{\sigma}_1^2+\beta_l^2\hat{\sigma}_2^2+\gamma^2\hat{\sigma}_3^2+2\alpha\beta\widehat{\textrm{GCov}}_A\right.\nonumber\\
&&\left.-2\alpha\gamma\widehat{\textrm{GCov}}_B+2\beta\gamma\widehat{\textrm{GCov}}_C\right].
\label {sufficient}
\end{eqnarray}

Eq (\ref{sufficient}) means that the only knowledge of the 6 elementary or final estimates is sufficient to compute the PDF of the actual set of $3M$  measurements. The number of estimates reduces to 2 in the absence of measurement noise (in this case, the final estimates can be computed from the elementary ones).

This result was expected: indeed, the variance estimates form a sufficient statistics for the variance estimation, meaning that the precise knowledge of the set of measurements $\vec{X}=(x_1, \ldots, x_n)^T$ does not bring any new information beyond the estimates. More precisely, the  vector of estimates $\vec{E}$ forms a sufficient statistics for $\vec{\Theta}$ if $p(\vec{X}|\vec{\Theta})=f(\vec{E},\vec{\Theta})\cdot g(\vec{X})$, where $f$ and $g$ are two functions \cite{saporta1990}. Indeed, we obtain after Bayesian inversion:
$p(\vec{\Theta}|\vec{X})\propto \pi(\vec{\Theta}) \cdot p(\vec{X}|\vec{\Theta})\propto  \pi(\vec{\Theta}) \cdot f(\vec{E},\vec{\Theta})\cdot g(\vec{X})$. In the experimenter world, $\vec{X}$ is formed by actual measurements, with known values. Hence, $g(\vec{X})$ appears as a constant. On the other hand, $p(\vec{\Theta}|\vec{X})$ is a function of the random variable $\vec{\Theta}$, whose integral is unity. Because of this normalisation, the multiplication by the constant $g(\vec{X})$ does not change the value of the function, that is entirely determined by $f(\vec{E},\vec{\Theta})$ and the prior.  However, it does not mean that only the PDF of the estimates makes sense. We prove in this paper that the best way is to use these estimates to compute the Gaussian PDF of the data themselves, that remains Gaussian after applying the K.L. transform.

Because we only use the estimates, the correlation between successive data does not lead to more complexity. Indeed, let $N$ be the number of degrees of freedom corresponding to a set of $M$ measurements, with $N<M$. The estimation of $N$ has been reported in \cite{greenhall}. We have simply to write Eq. (\ref{proba}) for $N$ independent measurements by using estimates computed on  $M$ correlated measurements, giving:

\begin{equation}
p(\vec{W}|\vec{\Theta})=\prod_{l=1}^{3}\frac{1}{V_l^{N/2}}\exp\left(-\frac{N\sum_{k=1}^M \sw_{l,k}^2}{2M V_l}\right)
\label{Nfreed}
\end{equation}

	\subsection{The KLTS algorithm\label{sec:KLTSalgo}}
To compute $p(\vec{\Theta}|\vec{W})$, we apply the same approach as in \cite{vernotte2018}. To take into account all the a priori values of $\vec{\Theta}$, we use a Monte-Carlo scheme with random sampling. This sampling ensures the observance of the total ignorance a priori law: the samples are chosen at random on a logarithmic scale, independently for each variance. We work in the experimenter point of view: we assume that two triplets{\tiny } of estimates, final and elementary, have been calculated from the $3 M$ elementary measurements. These six numbers have six definite values that will be used in the calculations detailed below. The different steps of the calculation are performed in the same order as in  \cite{vernotte2018} (the common steps of the two methods are recalled here, for sake of completeness):

\begin{itemize}
	\item Choose at random a triplet of true variances, with a uniform probability on a logarithmic scale for each variance and independence between the three variances.
	\item Calculate for this triplet the covariance matrix of the measurements given by Eq. (\ref {varab}) and (\ref{eq:theo_GCov}). 
	\item Calculate the eigenvectors and eigenvalues of this covariance matrix
	\item Use these values to calculate the PDF given in Eq. (\ref{Nfreed}), using Eq. (\ref{sufficient}).
	\item For an exhaustive exploration of the PDF, repeat $10^7$ times the entire process.
	\item For each of the three variables, normalize the probability densities by dividing by their sum (sum of  $10^7$ values) .
	\item Also for each of the three variables, sort the true variance values and calculate the cumulative density function by a partial sum on the associated normalized probability densities. 
	\item determine a confidence interval at $95 \%$ on each true variance from the corresponding cumulative density functions.
	\item Verify that the low limit of the confidence interval is meaningful. For a Gaussian distribution, $99.7 \%$ of data are included in a confidence interval at $\pm 3 \sigma$. If the low limit of this $\pm 3 \sigma$ confidence interval (in logarithmic scale) is smaller than the low limit of the a priori range (here $10^{-5}$ ), we suspect (and have verified)  that the low limit of the $\pm 2\sigma$ confidence interval calculated in the preceding step will depend on the low limit of the a priori range. If it occurs, we replace the low limit of the confidence interval by 0.
\end{itemize}

Finally, we calculate the median value, i.e. the argument giving the CDF equal to 0.5. This value, always positive, may be an alternative estimate of the parameters.

\section{Validation of the KLTS Method by Monte-Carlo Simulations}
	\subsection{Principle of the simulation}
In order to validate the KLTS method, we have compared its results to Monte-Carlo simulations. 
%

The algorithm is as follows: \label{sec:algo}
\begin{enumerate}
	\item Select a target set of $3$ final estimates $(\hat{\sigma}_A^2, \hat{\sigma}_B^2, \hat{\sigma}_C^2)=(A_0,B_0,C_0)$. We call it ``reference estimate set''.
	\item Draw at random a parameter triplets $(\sigma_A^2, \sigma_B^2, \sigma_C^2)$.
	\item Randomly draw $3N$ measurements $\{\sz_{ABk},\sz_{BCk},\sz_{CAk}\}$ according to $(\sigma_A^2, \sigma_B^2, \sigma_C^2)$. 
	\item Compute the final estimates $\left(\widehat{\textrm{GCov}}_A, \widehat{\textrm{GCov}}_B, \widehat{\textrm{GCov}}_C\right)$ from these measurements.
	\item If this final estimate set is close to the reference estimate set within 10 \%\footnote{This 10 \% interval was selected by a compromise ensuring that the final estimate set is sufficiently close on a log scale to the reference estimate set, while the calculation time is reasonable (the computation time increases as the cube of the inverse of the interval width).}, the corresponding parameter triplet $(\sigma_A^2, \sigma_B^2, \sigma_C^2)$ is kept, otherwise it is thrown.
	\item Go to Step 2).
\end{enumerate}
Each simulation run stops when 10,000 achievements have been obtained in order to provide a meaningful knowledge of the parameter statistical distributions.

This ensemble of 10,000 parameter triplets giving $(\hat{\sigma}_A^2, \hat{\sigma}_B^2, \hat{\sigma}_C^2)=(A_0,B_0,C_0)$ is then compared to the confidence interval obtained by the KLTS method.

Thanks to the sufficient statistic properties of the estimates, it turns out that any $3N$ measurement set providing the reference estimate set leads to the same statistical distribution of the parameter triplet.

	\subsection{Results and discussion}
   
We will focus this study only on the efficiency of the method, i.e. its ability to fit the true confidence intervals, and not on the behavior of the confidence intervals versus different parameters. More details on this latter subject may be found in \cite{vernotte2018}. 

This validation way checks the accuracy of the KLTS method with respect to several physical variables.

		\subsubsection{Influence of the EDF}
\begin{figure}
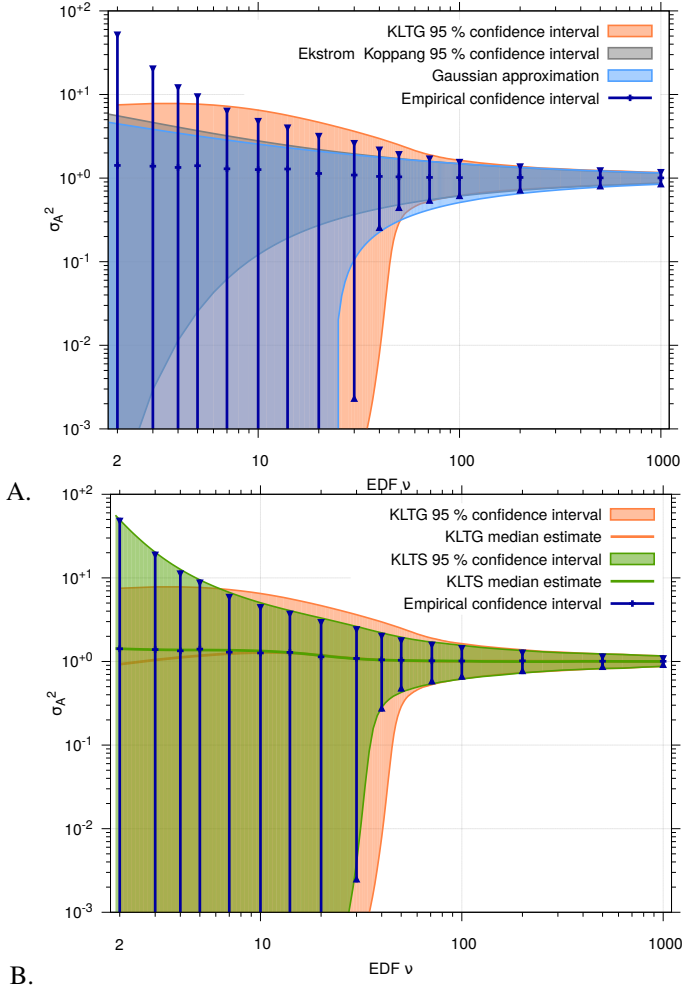

\hspace{-0.5mm}A.
\vspace{-2mm}
\includegraphics[width=\columnwidth]{A1_vs_nu_unc_log_KLTG_3.pdf}
\hspace{-0.5mm}B.
\vspace{-2mm}
\includegraphics[width=\columnwidth]{A1_vs_nu_unc_log_KLTS.pdf}
\caption{Estimation of the 95 \% confidence interval for a set of clocks with equal final estimates ($\hat{\sigma}_A^2=\hat{\sigma}_B^2=\hat{\sigma}_C^2=1$) versus the EDF number. The graph above (A) shows the results obtained from previous methods and the graph below (B) presents the results of the KLTS method (green area) compared to the KLTG (red area). The reference is given by the blue error bars which were obtained by massive Monte-Carlo simulations (see algorithm \S \ref{sec:algo}).\label{fig:vsEDF}} 
\end{figure}

In order to compare the confidence interval given by several other methods, we chose to set the final estimate values to $\hat\sigma_A^2=\hat\sigma_B^2=\hat\sigma_C^2=1$ and to vary the number of EDF. The results given by the different methods, a simple Gaussian approximation \cite{vernotte2016}, the Ekstrom-Koppang method (EK) \cite{ekstrom2006} and the KLTG method \cite{vernotte2018} are plotted in Figure \ref{fig:vsEDF}-A. All these confidence intervals fit pretty well the empirical error bars obtained by 10,000 Monte-Carlo simulations above 200 EDF. Between 50 and 200 EDF, EK and KLTG remain  usable but only KLTG fits quite well between 5 and 50 EDF. Finally, below 5 EDF, none of these methods is satisfactory.

The new KLTS method is compared to KLTG in Figure \ref{fig:vsEDF}-B. The KLTG confidence intervals fit pretty well the empirical error bars even for 2 EDF. The same is true for the median estimates which correspond perfectly to the empirical median values at the center of the error bars. 

\begin{table}
\caption{Comparison for 1 EDF of the confidence intervals as well as the median estimates (50 \%) obtained by the Monte-Carlo simulations (Empirical), by the KLTG method of \cite{vernotte2018} and by the new KLTS method. The $(A_0,B_0,C_0)$ estimate triplet is $(-0.5, 1, 1)$.
}
\label{tab:inverse}
\begin{center}
\begin{tabular}{lc|cc|cc}
 & Bounds & \multicolumn{2}{c|}{$\hat{\sigma}_A^2=-0.5$} & \multicolumn{2}{c}{$\hat{\sigma}_B^2=\hat{\sigma}_C^2=1$} \\ 
\hline
Emp. & 2.5 \% & $1.78\cdot 10^{-5}$ & (2.5 \%) & $3.0\cdot 10^{-5}$ & (2.5 \%)\\
& 50 \% & 0.192 & (50.0 \%) & 0.90 & (50.0 \%)\\
& 95 \% & 27.5 & (95.0 \%) & 76 & (95.0 \%)\\
& 97.5 \% & 92 & (97.5 \%) & 182 & (97.5 \%)\\
\hline
KLTG & 2.5 \% & $1.36\cdot 10^{-5}$ & (1.52 \%) & $1.23\cdot 10^{-4}$ & (5.31 \%)\\
& 50 \% & $5.5\cdot 10^{-3}$ & (28.7 \%) & 0.42 & (37.0 \%)\\
& 95 \% & 0.83 & (68.8 \%) & 3.23 & (70.4 \%)\\
& 97.5 \% & 1.38 & (75.0 \%) & 5.3 & (76.3 \%)\\
\hline
KLTS & 2.5 \% & $1.67\cdot 10^{-5}$ & (2.59 \%) & $2.86\cdot 10^{-5}$ & (2.37 \%)\\
& 50 \% & 0.200 & (50.6 \%) & 0.90 & (49.8 \%)\\
& 95 \% & 35 & (95.5 \%) & 90 & (95.3 \%)\\
& 97.5 \% & 98 & (97.8 \%) & 208 & (97.6 \%)\\
\end{tabular}
\end{center}
\end{table}

However, this comparison is limited to 2 EDF because the set $\hat\sigma_A^2=\hat\sigma_B^2=\hat\sigma_C^2=1$ is impossible for 1 EDF without measurement noise. It was demonstrated that, if the measurement noise is negligible, and this is always the case when there is only 1 EDF, the 3$^\textrm{\footnotesize rd}$ final estimate is totally determined by the other 2 ones \cite{vernotte2018}. For this reason, we set $\hat{\sigma}_B^2=\hat{\sigma}_C^2=1$ which leads to $\hat{\sigma}_A^2=-0.5$. Table \ref{tab:inverse} compares 
the confidence interval obtained by KLTG and KLTS to the empirical bounds given by 10,000 Monte-Carlo simulations. 

Unlike the results of KLTG, the KLTS bounds as well as the median estimate show a good agreement with the empirical bounds and median. However, because of the very low level of the 2.5 \% bounds, they should be considered as equal to 0. In these conditions, KLTS gives fully reliable 95 \% upper limits (95.7 \% for $\sigma_A^2$, 95.6 \% for $\sigma_B^2$ and  $\sigma_C^2$). 

		\subsubsection{Influence of the disparity of the 3 final estimates}
\begin{figure}
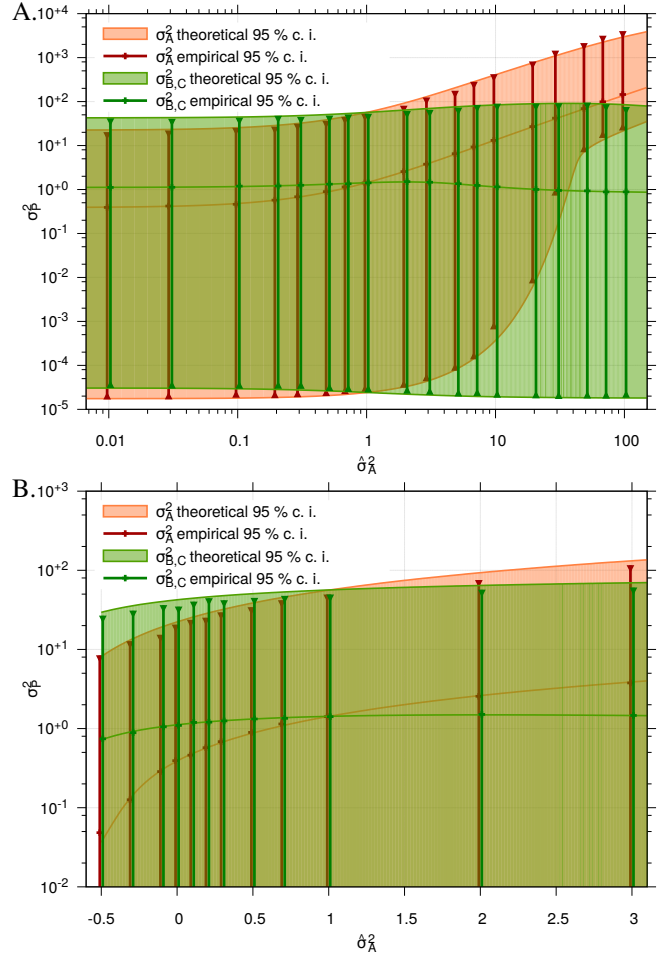

\hspace{-0.5mm}A.

\vspace{-2mm}

\includegraphics[width=\columnwidth]{nu2_ABC_unc_log.pdf}

\vspace{-5mm}

\hspace{-0.5mm}B.

\vspace{-2mm}

\includegraphics[width=\columnwidth]{nu2_ABC_unc_lin.pdf}
\caption{Estimation of the 95 \% confidence intervals for 2 clocks with equal final estimates ($\hat{\sigma}_B^2=\hat{\sigma}_C^2=1$) versus the final estimate of the other clock ($-0.5\leq\hat{\sigma}_A^2\leq 100$). The number of EDF is 2. The graph above (A) shows the results obtained from positive final estimates on a log-log plot whereas the graph below (B) uses a linear scale in order to display also the results obtained from negative final estimates. The error bars, red for $\sigma_A^2$ and green for $\sigma_B^2$ and $\sigma_C^2$, were obtained by massive Monte-Carlo simulations.\label{fig:vssA2}} 
\end{figure}

Figure \ref{fig:vssA2} displays, for 2 EDF, the evolution of the confidence interval of all parameters versus the final estimates $\hat\sigma_A^2$ which varies from 0.01 to 100 whereas the 2 other ones are set to $\hat\sigma_B^2=\hat\sigma_C^2=1$. 
 On the log-log plot as well as on the linear plot, the agreement between the error bars and the KLTS confidence intervals is excellent.

		\subsubsection{Influence of the measurement noise}
\begin{figure}
\centering
\includegraphics[width=\columnwidth]{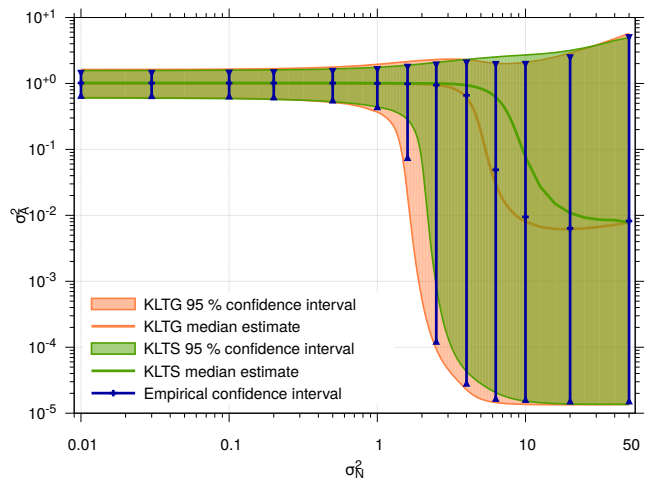}
\caption{Estimation of the 95 \% confidence intervals for a set of clocks with equal noise measurements ($\hat{\sigma}_A^2=\hat{\sigma}_B^2=\hat{\sigma}_C^2=1$) versus the measurement noise level ($0.01\leq\hat{\sigma}_N^2\leq 50$). The number of EDF is 100. The results obtained by the KLTS method (green area) may be compared to the ones obtained by KLTG (red area). The blue error bars were obtained by massive Monte-Carlo simulations.\label{fig:vssN2}} 
\end{figure}

The main advantage of GCov over the three-cornered hat relies in its rejection of the measurement noise. Since the measurement noise is directly addressed by KLTS, e.g. to solve equation (\ref{sufficient}), it is of importance to study its influence on the confidence intervals. This influence is shown on Figure \ref{fig:vssN2} where the final estimates are set to ($\hat{\sigma}_A^2=\hat{\sigma}_B^2=\hat{\sigma}_C^2=1$), the number of EDF is 100 and  the variance of the measurement noise $\sigma_N^2$ varies from 0.01 to 50. Despite slight discrepancies between the confidence intervals and the error bars obtained from the Monte-Carlo simulations, the agreement is quite good. The main differences appear for $1<\sigma_N^2<10$ when the 2.5 \% bound decreases drastically. This effect is also visible for the median estimate which seems to decrease at a different rate for the theoretical and the empirical median estimates. A very small discrepancy is also visible for the upper bound at $\sigma_N^2 = 7$, 10 and 20. These discrepancies can be explained as follows: in the PDF computation of KLTS, the product (\ref{Nfreed}) of the elementary probabilities of each data becomes extremely small when the number of data becomes large, and the result of (\ref{Nfreed}) can be  wrongly truncated to 0 even in double precision. As a consequence, the use of KLTS must be restricted for cases where the number of EDFs does not exceed one hundred. For hundred EDfs as in Figure \ref{fig:vssN2}, the truncation errors already explain some discrepancies for a high level of measurement noise, that do not however affect much the upper bound of the confidence interval, undoubtedly the most important result.

		\subsubsection{Discussion: KLTS vs KLTG}
KLTS is undoubtedly a rigorous approach which does not rely on approximations or simplifying assumptions. As a consequence, it provides very relevant confidence intervals even for very low EDF, including the limit case of 1 EDF (see Table \ref{tab:inverse}), as proved by the almost perfect agreement between the confidence intervals given by KLTS and obtained by the Monte Carlo simulations in Figures \ref{fig:vsEDF} to \ref{fig:vssN2}. 
 The only slight differences can be attributed to the way of computing the PDF by numerical integration.

However, we have seen that truncation errors can occur with  KLTS when the number of data becomes large: the use of KLTS must be restricted for cases where the number of EDFs does not exceed one hundred.

On the other hand, KLTG relies on a Gaussian approximation of the estimates \cite{vernotte2018} which can be assumed only for high EDF. Unlike KLTS, KLTG must be restricted for high EDFs. Figure \ref{fig:vssN2} shows that KLTG and KLTS provide almost the same results for 100 EDFs and these results are in perfect agreement with the Monte-Carlo simulations. KLTG is then a very good substitute to KLTS above 100 EDFs.

	\subsection{Application to a set of real clocks}
\begin{figure}
\centering
\includegraphics[width=\columnwidth]{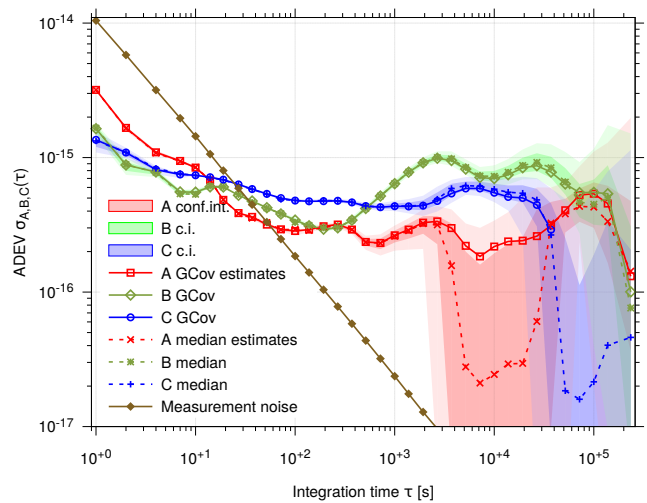}
\caption{GCov measurements of a set of 3 Cryogenic Sapphire Oscillators of FEMTO-ST. The confidence intervals are represented by colored areas: bright colors for 68 \% c.i. (1 $\sigma$) and pale colors for 95 \% c.i. (2 $\sigma$). The brown line shows the measurement noise obtained from the closure. The linear frequency drifts of the CSO have been removed.\label{fig:CSO}} 
\end{figure}

These methods were applied to assess the confidence intervals for the GCov measurements of a set of 3 Cryogenic Sapphire Oscillators designed and made in FEMTO-ST. These measurements were carried out by the Tracking DDS designed and made in INRIM (for more details, see \cite{calosso2018}). It may be noticed that, since the clocks are in the same room, we can not fully assume their total independence. Figure \ref{fig:CSO} presents the same measurements as the ones plotted in Figure 11 (below) of \cite{calosso2018} but we have added the 95 \% confidence intervals obtained by using KLTS (below 100 EDF, i.e. $\tau> 5000$ s) and KLTG (above 100 EDF, i.e. $\tau< 5000$ s).

As expected, the confidence intervals are pretty tight around the estimates for low $\tau$ values whereas they extend downward for high $\tau$ values. The lower bounds tend toward 0 above $\tau=2\,000$ s for CSO A, above $\tau=20\,000$ s for CSO C and above $\tau=100\,000$ s for CSO B. In these cases, only an upper limit of the stability of the clock may be assessed. As expected, the less stable clocks have the smallest confidence intervals and we can see these intervals increasing or decreasing regarding the relative positions of the other CSOs. For instance, A is the most stable clock between 500 s and 30\,000 s but becomes the less stable clock around $10^5$ s, as a consequence, the lower bound of its confidence interval increases 
 from $\sim 0$ at $10^4$ s to $\sim 2\cdot 10^{-16}$ at $10^5$ s.

The median estimates remain generally close to the GCov estimates. However, when the lower bound of a clock tends toward 0, the median estimate decreases significantly and becomes far lower than the GCov estimate when this latter exists, i. e. when it is positive. Nevertheless, there is still a positive median estimate even when the GCov estimate is negative (e.g. C above $\tau=30\,000$ s). 
In such a case the upper limit of the confidence interval is
the relevant information, while the estimate is of little value.
It may be noticed that the relevance of the median estimates could be improved by using a more stringent prior $\pi(\vec\Theta)$ in Eq. (\ref{eq:bayes_inference}), i.e. a prior based on a perfectly objective \textit{a priori} knowledge of the range in which the parameters can vary. To stay in the most general case, we have preferred to stick to a ``total ignorance'' prior (see \ref{sec:KLTSalgo}) in this paper.

We have also added the ADEV of the measurement noise on Figure \ref{fig:CSO}. It is prevailing below 10 s and we can see that the confidence interval of the most stable clock in this range, B and C, have a larger confidence interval (e.g. C at 1 s) than for $\tau=100$ s for instance. At 1 s, the measurement noise is approximately 5 times higher than the stability of C and, despite its huge EDF number (EDF = 370\,000 @ 1 s), the effect of the measurement noise affects clearly its confidence interval as in Figure \ref{fig:vssN2}. In other words, the measurement noise is too high to be fully rejected by GCov.

\section{Conclusion}
The method we propose, KLTS, provides the Cumulative
Density Functions of the clock noise variances and the corresponding confidence intervals. Unlike the direct GCov estimates, the medians of these intervals cannot be negative.
 
KLTS is a rigorous method since it involves neither approximations nor simplifying assumptions. It is based on the property of sufficient statistics that form the estimates.

KLTS is valid for very low EDF including 1 EDF. Massive Monte-Carlo simulations have perfectly validated KLTS in different contexts: confidence interval versus the Equivalent Degrees of Freedom, vs the the stability disparity of the clocks and vs the level of the measurement noise.

However, KLTS suffers from the drawback of not being easily computed for high EDF ($>100$). But it has been proved in this paper that the KLTG method, presented in a previous paper \cite{vernotte2018}, is perfectly reliable above 100 EDF even if the measurement noise is strong. 
The combination of these two methods provides then a powerful tool to assess confidence intervals for the clock noise variances, whatever the EDF and the measurement noise level (see \cite{sigmatheta} for a software solution). 



\section*{Acknowledgement}

This work was partially funded by the ANR Programmes
d'Investissement d'Avenir (PIA) Oscillator IMP (Project 11-EQPX-0033) and FIRST-TF (Project 10-LABX-0048) as well as by the Conseil Régional de Bourgogne Franche-Comté.


\bibliography{3CH_uncertainties.bib}

\section*{APPENDIX: Glossary of symbols}
\subsection{Measurements}
\begin{tabular}{lp{7cm}}
\hline
$\bar{\sy}_{ABk}$ &  $k^\textrm{\footnotesize th}$ frequency deviation sample between clock $A$ and clock $B$\\
\hline
$\bar{\sy}_{1k}$ &  $k^\textrm{\footnotesize th}$ frequency deviation sample at the output of counter $1$\\
\hline
$\bar{\sy}_{N1k}$ &  $k^\textrm{\footnotesize th}$ frequency deviation sample of the intrinsic measurement noise of counter $1$\\
\hline
${\sz}_{ABk}$ &  Difference between the two consecutive clock frequency deviation samples $\bar{\sy}_{ABk+1}$ and $\bar{\sy}_{ABk}$ divided by $\sqrt{2}$\\
\hline
${\sz}_{1k}$ &  Difference between the two consecutive counter frequency deviation samples $\bar{\sy}_{1k+1}$ and $\bar{\sy}_{1k}$ divided by $\sqrt{2}$\\
\hline
${\sz}_{N1k}$ &  Difference between the two consecutive noise frequency deviation samples $\bar{\sy}_{N1k+1}$ and $\bar{\sy}_{N1k}$ divided by $\sqrt{2}$\\
\hline
${\sz}_{cls,k}$ &  Sum of the $k^\textrm{\footnotesize th}$ $\sz$ measurements of all counters (closure relationship)\\
\hline
\end{tabular}

\subsection{Parameters and estimates}
The list below gives the parameters, i.e. the mathematical expectation we want to assess. The same symbols with a hat $\hat{\cdot}$ stand for the estimates of these parameters, i.e. the values computed from a finite number of measurements.

\begin{tabular}{lp{6.5cm}}
\hline
$\sigma_{AB}^2$ &  Allan variance of the comparison between clock $A$ and clock $B$\\
\hline
$\sigma_{1}^2$ &  Allan variance of the output of counter $1$ (\textbf{elementary estimate})\\
\hline
$\sigma_{N1}^2$ &  Allan variance of the intrinsic measurement noise of counter $1$. Since the noise level is assumed to be the same for all counters, the subscript $1$, $2$ or $3$ may be omitted.\\
\hline
$\sigma_{cls}^2$ &  Allan variance of the closure\\
\hline
$\sigma_{A}^2$ &  Allan variance of clock $A$ ($\sigma_{A}^2$ is called \textbf{model parameter} whereas $\hat\sigma_{A}^2$ is called \textbf{final estimate})\\
\hline
$\textrm{TCH}_A$ &  Three-cornered hat computation of Allan variance for clock $A$\\
\hline
$\textrm{GCov}_A$ &  Groslambert covariance computation of Allan variance for clock $A$\\
\hline
\end{tabular}

\subsection{Bayesian formalism}
\begin{tabular}{lp{6.3cm}}
\hline
$p\left(\vec{X}|\vec{\Theta}\right)$ &  Conditional probability density function of the multidimensional measurement $\vec{X}$ knowing the multidimensional parameter $\vec{\Theta}$ (model world)\\
\hline
$p\left(\vec{\Theta}|\vec{X}\right)$ &  Conditional probability density function of the multidimensional parameter $\vec{\Theta}$ knowing the multidimensional measurement $\vec{X}$ (experimenter world)\\
\hline
$\pi\left(\vec{\Theta}\right)$ &  \textit{A priori} probability density function of the multidimensional parameter $\vec{\Theta}$ before any measurement (prior)\\
\hline
\end{tabular}

\subsection{KLTS method}
\begin{tabular}{lp{4.5cm}}
\hline
$\left(\begin{array}{lcr}
\alpha_1 & \beta_1 & \gamma_1\\
\alpha_2 & \beta_2 & \gamma_2\\
\alpha_3 & \beta_3 & \gamma_3
\end{array}\right)$ & Eigenvectors of the rotation matrix which diagonalizes
the covariance matrix of the measurements (Karhunen-Lo\`eve Transform)\\
\hline
$\vec{\Theta}=(V_{1}, V_{2}, V_{3})^T$ & KLT variances, i.e. eigenvalues of the covariance matrix of the measurements\\
\hline
$\vec{W}=(\sw_{1k}, \sw_{2k}, \sw_{3k})^T$ & Image of the measurements $(\sz_{1k}, \sz_{2k}, \sz_{3k})^T$ by the KLT\\
\hline
$M$ & Number of measurements\\
\hline
$N$ & Number of Equivalent degrees of freedom\\
\hline
\end{tabular}

\end{document}